\documentclass[pra,twocolumn,showpacs,preprintnumbers,amsmath,amssymb]{revtex4}
\usepackage{graphicx}
\usepackage{dcolumn}
\usepackage{bm}
\usepackage{stmaryrd}
\usepackage{latexsym}
\usepackage{amssymb}
\usepackage{amsfonts}
\usepackage{amsmath}

\begin{document}
\title{Double occupancies in confined attractive fermions on optical lattices}
\author{Ji-Hong Hu}
\affiliation{Department of Physics, Zhejiang Normal University,
Jinhua 340012, China}
\author{Gao Xianlong}
\email{gaoxl@zjnu.edu.cn}
\affiliation{Department of Physics, Zhejiang Normal University,
Jinhua 340012, China}

\date{\today}
\begin{abstract}
We perform a numerical study of a one-dimensional Fermion-Hubbard model in harmonic traps within the Thomas-Fermi approximation based on the exact Bethe-ansatz solution. The $\rho-U/t$ phase diagram is shown for the systems of attractive interactions ($\rho$ is the characteristic density and $U/t$ the interaction strength scaled in units of the hopping parameter.). We study the double occupancy, the local central density and their derivatives. Their roles are discussed in details in detecting the composite phases induced by the trapping potential.
\hspace*{7.0cm}
\end{abstract}
\maketitle
The recent remarkable experimental progress in cooling and manipulating ultracold atomic gases in optical lattices
provides us an alternative way to investigate the many-body effects in condensed matter system~\cite{review paper}.
For the optical lattices loaded with atomic gases, the on-site interaction between different species can be tuned to be very strong by means of a technique named Feshbach resonance~\cite{Feshbach} or by increasing the lattice depth to decrease the hopping between neighbor sites. Thus the outstanding controllability offers the possibility to use these systems to "quantum simulate" quantum many-body physics in a well-controlled way. An example is the experimentally realized Tonks-Girardeau gas in a one-dimensional (1D) optical lattice~\cite{Paredes}.

In the experiments, the trapping potential used to confine atoms induces inhomogeneity and complexity in characterizing the quantum
phases. Though a few techniques exist in experiments observing the new quantum phases and probing various properties in ultracold gases in optical lattices, such as the time-of-flight imaging of the momentum distribution, noise correlations and Bragg spectroscopy~\cite{review paper}, there are difficulties in observing the composite phases induced by the inherent trapping potential, for example, the incompressible bulk Mott- and band-insulating phases surrounded by the compressible fluid phases at the edges in the 1D Fermi-Hubbard model under harmonic confinement. Scarola et al. proposed to use the core compressibility by filtering out the edge effects, which offers a direct probe of incompressible phases independent of inhomogeneity~\cite{Scarola}. Duan et al. suggested to use the Fourier sampling of time-of-flight images to reveal new correlation functions~\cite{Duan}. Kollath et al. used the double occupancy (DO) induced by periodic lattice modulations via the creation of molecules to identify the pair-binding energy and the spin-ordering in a Fermi gas in an optical lattice~\cite{Kollath}, which was recently measured in the experiments by molecule formation and used to identify an incompressible Mott-insulator phase of two hyperfine states of strongly repulsive fermionic $^{40}$K atoms~\cite{Jordens}. The compressibility of the quantum gas in the trap was used recently in identifying the incompressible Mott-insulating phase at strong interactions by independent control of external confinement and lattice depth~\cite{Schneider}. Here in this paper we discuss how to characterize the different composite phases of 1D confined fermions on optical lattices, by investigating observables like the double occupation sites, the local central density and their derivatives.

We consider a two-component Fermi gas in a tube with $N_f$ atoms and $N_s$ lattice sites, which can be realized by a strong confinement in the transverse directions~\cite{moritz_2005} with an additional periodic potential applied along the tube and can be described by a one-band inhomogeneous Fermi-Hubbard model~\cite{jaksch_98},
\begin{eqnarray}\label{eq:hubbard}
\hat {H}_s&=&-t\sum_{i=1,\sigma}^{N_s}({\hat
c}^{\dagger}_{i\sigma}{\hat c}_{i+1\sigma}+{H}.{c}.)+
U\sum_{i=1}^{N_s}\,{\hat D}_{i}\nonumber\\
&&+V_2\sum_{i=1}^{N_s}(i-N_{s}/2)^2{\hat n}_{i}
\end{eqnarray}
where the spin degrees of freedom $\sigma=\uparrow,\downarrow$ are pseudospin-$1/2$ labels for
two internal hyperfine states. ${\hat c}_{i\sigma}$ (${\hat c}^{\dagger}_{i\sigma}$) are fermionic operators annihilating (creating) particles with spin $\sigma$ in a Wannier state at site $i$. ${\hat n}_i= \sum_{\sigma} {\hat n}_{i\sigma}=\sum_{\sigma} {\hat c}^{\dagger}_{i\sigma}{\hat c}_{i\sigma}$ is the total site occupation operator, $t$ is
the tunneling between nearest neighbors, $U$ is the strength of the on-site attractive interaction, ${\hat D}_{i}={\hat n}_{i\uparrow}{\hat n}_{i\downarrow}$ is the double occupancy operator at site $i$, and $V_i$ describes the strength of the trapping potential.

The homogeneous 1D Fermi-Hubbard model belongs to the universality class of Luttinger liquids. At zero temperature, the properties of this model in the thermodynamic limit ($N_{\sigma}, N_s\rightarrow \infty$, but with finite $N_{\sigma}/N_s$) are determined by the fillings $n_\sigma=N_\sigma/N_s$ and by the dimensionless coupling constant $u=U/t$. According to Lieb and Wu~\cite{LiebWu}, the ground state properties for different fillings in the thermodynamic limit are described by the coupled integral equations (details see Refs. \onlinecite{KocharianPRB,gaoprb78}).
\begin{figure}
\begin{center}
\includegraphics*[width=0.9\linewidth]{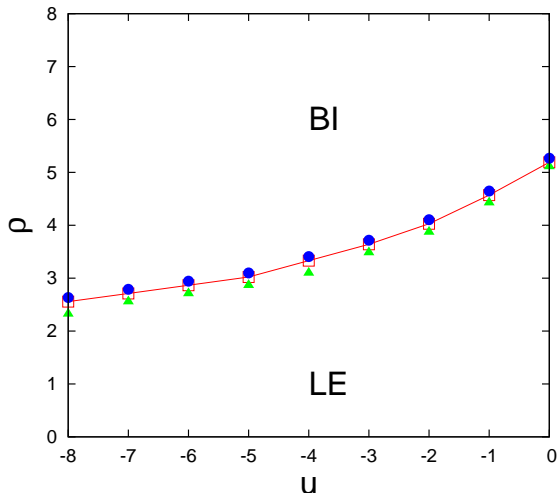}
\caption{(Color online) Phase diagram for the one-dimensional inhomogeneous fermion-Hubbard model as a function of attractive interaction strength $u=U/t$ and characteristic particle density $\rho=N_f\sqrt{V_2/t}$. For fixed $u$, the shape of the density profile is completely determined by $\rho$. The band insulator (BI) phase is determined with the local occupation in the trap center satisfying $|n_i-2|<0.005$. Below the BI phase, it is the Luther-Emery (LE) liquid phase. The boundary between BI and LE phases is plotted as open squares ($\square$) with solid lines as a guide for the eye. The solid circles ($\bullet$) represent the phase boundary determined from the derivative of the double occupancy, and the solid upper triangles ($\blacktriangle$) represent the one determined from the derivative of the local central density (See details in the text and in Fig. \ref{fig:four}). All of them can be used as identifying the onset of the BI phase.
\label{fig:one}}
\end{center}
\end{figure}
\begin{figure}
\begin{center}
\includegraphics*[width=0.9\linewidth]{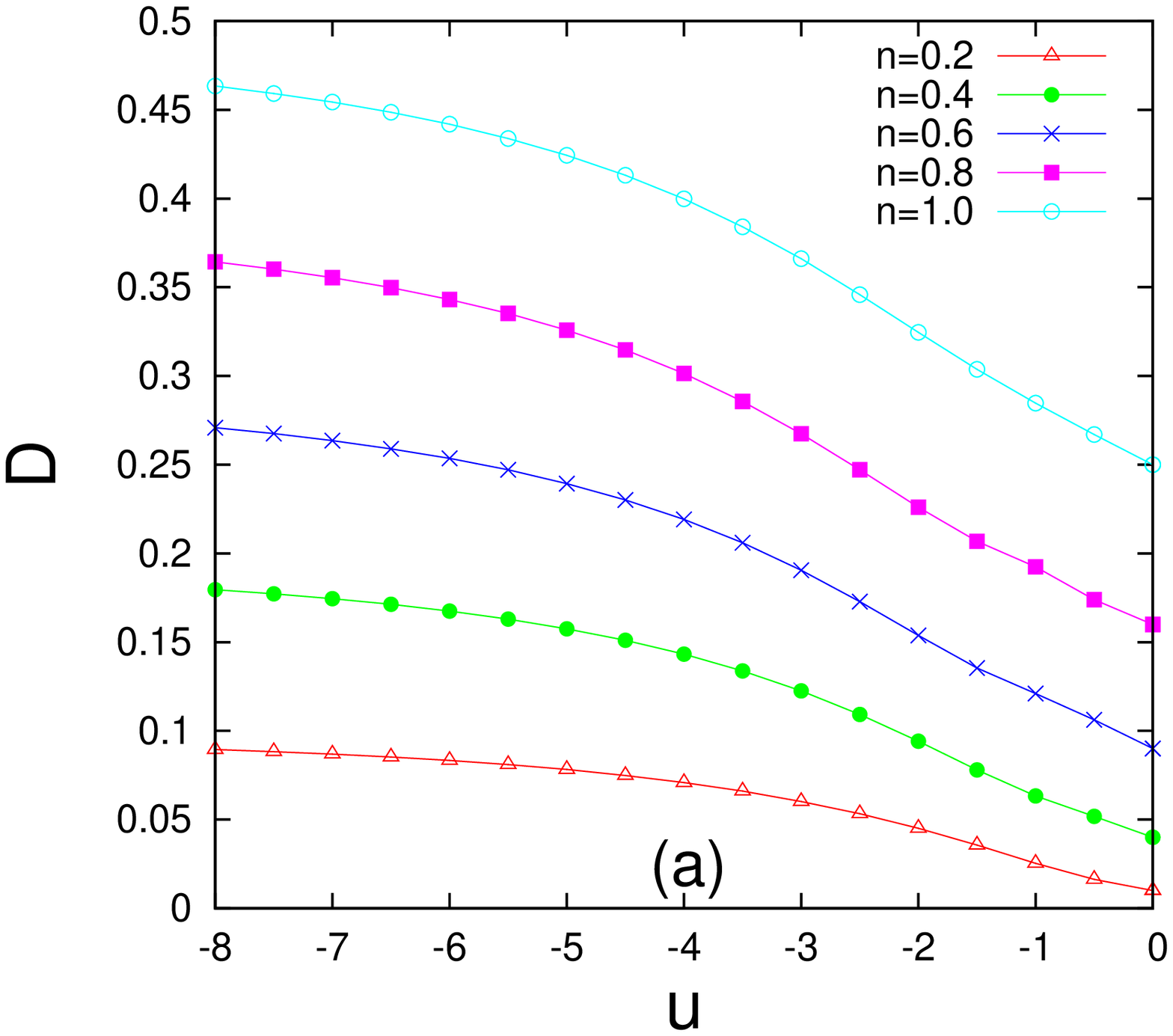}\\
\includegraphics*[width=0.9\linewidth]{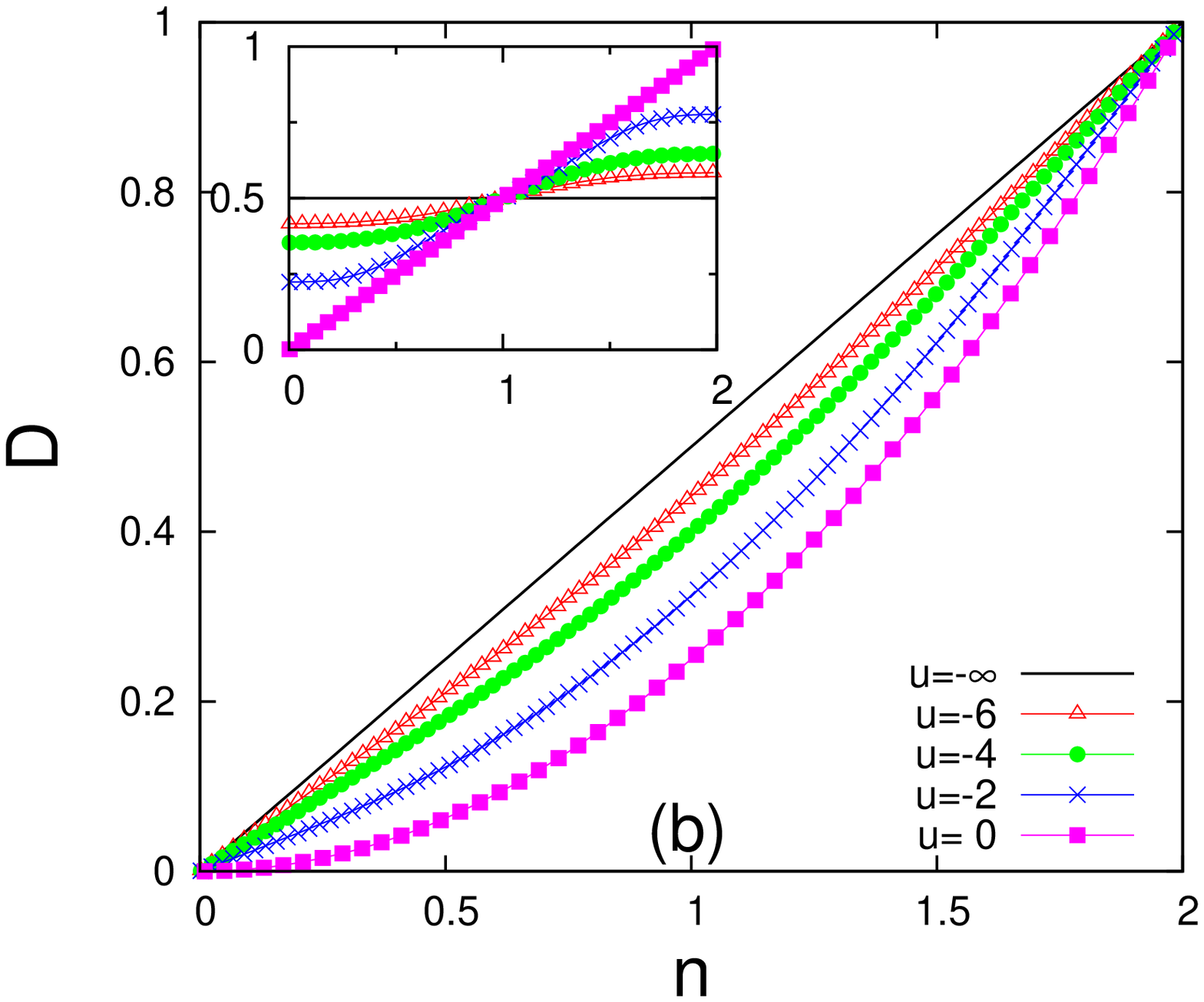}
\caption{(Color online) The ground-state concentration of doubly occupied sites in the homogeneous system $D^{\rm hom}$ (a) as a function of interaction strength $u$ for various $n$ and (b) as a function of particle density $n$ for various interacting strengths $u$. For $0\le n\le 2$, $D^{\rm hom}$ is in the range $0\le D^{\rm hom}\le 1$. For fixed $n$, $D^{\rm hom}$ increases monotonically with increasing $|u|$ and for fixed $u$ it increases monotonically with increasing $n$. At $U=0$, $D^{\rm hom}=n^2/4$ and at strongly attractive coupling limit $D^{\rm hom}= n/2$. The attractive interaction favors the formation of pairs between opposite spins. In the inset of (b), the derivative of the double occupancy with respect to the particle density $n$ is plotted, which is used to calculate the double occupancy for the inhomogeneous system in the spirit of local density approximation.
\label{fig:two}}
\end{center}
\end{figure}

The inhomogeneity of 1D Fermi-Hubbard model caused by the boundary conditions or the external potential,
normally invalidates a reliable analytical method that usually used in the homogeneous system.
Many numerically accurate schemes, such as the density-matrix renormalization group~\cite{Machida1,gaoprb78,Molina}, quantum-Monto-carlo (QMC)~\cite{Rigol,Rigol2004,Astrakhardik,Casula}, exact diagnolization~\cite{Machida1,exact diagnolization,Nikkarila} and density-functional theory based on the exact Bethe-ansatz solution~\cite{gaoprb78}, are used in obtaining the phase diagram and the collective oscillations of the atomic mass density. The Thomas-Fermi approximation based on the exact Bethe-ansatz solution has also been successfully used in characterizing the phase diagrams~\cite{Liu}. The rich phases induced by the trapping potential or the phase transitions driven by the system parameters can be identified by the site occupation, the variance of the local density, the local compressibility~\cite{Rigol2004}, or the thermodynamic stiffness (i.e., the inverse of the global compressibility)~\cite{gaoprb78}.

Here we study the inhomogeneous Fermi-Hubbard model described by Eq. (\ref{eq:hubbard}),
which allows for the spatial coexistence of different quantum states~\cite{Rigol,Liu}.
For attractive interactions, $U<0$, there are two different composite phases, well classified in the phase diagram
defined by a scaled dimensionless characteristic density $\rho=N_f\sqrt{V_2/t}$ and $u=U/t$~\cite{Rigol}: the Luther-Emery (LE) liquid phase characterized by fermion pairs and atomic density wave, and band insulator (BI) phase with a plateau of $n_i=2$ in the trap center of tightly bound spin-singlet dimers surrounded by Luther-Emery layers~\cite{gaoprb78}.

All these phases have been analyzed in the $\rho-u$~\cite{Rigol2004,Heiselberg}, $\mu-u$ ($\mu$ is the chemical potential)~\cite{Campo}, and $n_0-\nu$~\cite{Liu} ($n_0$ is the central density and $\nu=2\sqrt{N_f}/\pi$ is the characteristic filling factor) phase diagrams.

\begin{figure}
\begin{center}
\tabcolsep=0 cm
\begin{tabular}{cc}
\scalebox{0.25}[0.25]{\includegraphics{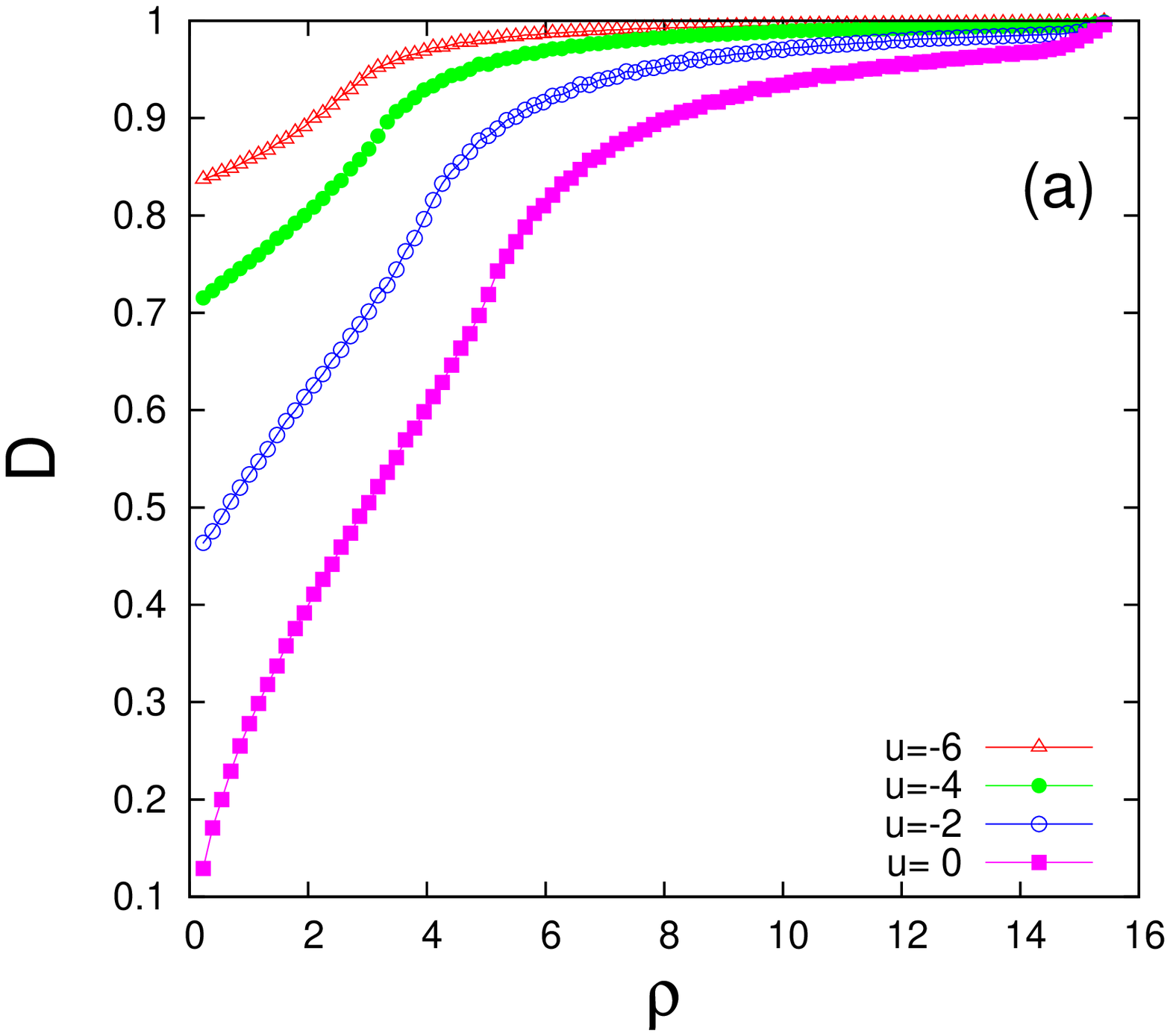}}&
\scalebox{0.25}[0.25]{\includegraphics{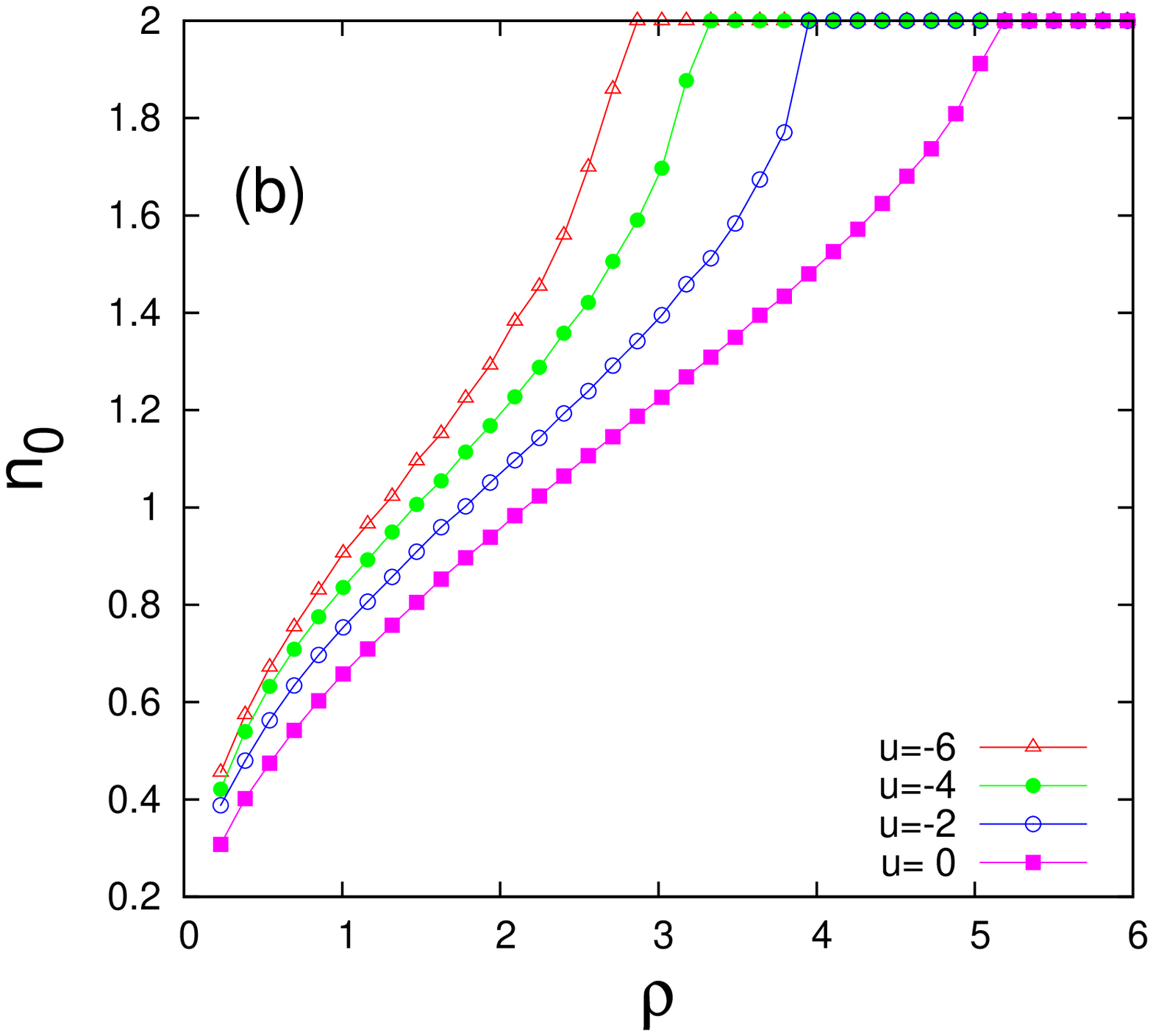}}\\
\end{tabular}
\caption{(Color online) (a) The ground-state concentration of doubly occupied sites $D$ as a function of characteristic density $\rho$ in the system under the presence of trapping potential for various interacting strength $u$. (b) The local central density $n_0$ as a function of characteristic density $\rho$ for various interacting strength $u$.
For comparison, the double occupancy and the local central density for the noninteracting case ($u=0$) are also shown in (a) and (b), respectively.
\label{fig:three}}
\end{center}
\end{figure}
\begin{figure}
\begin{center}
\includegraphics*[width=0.9\linewidth]{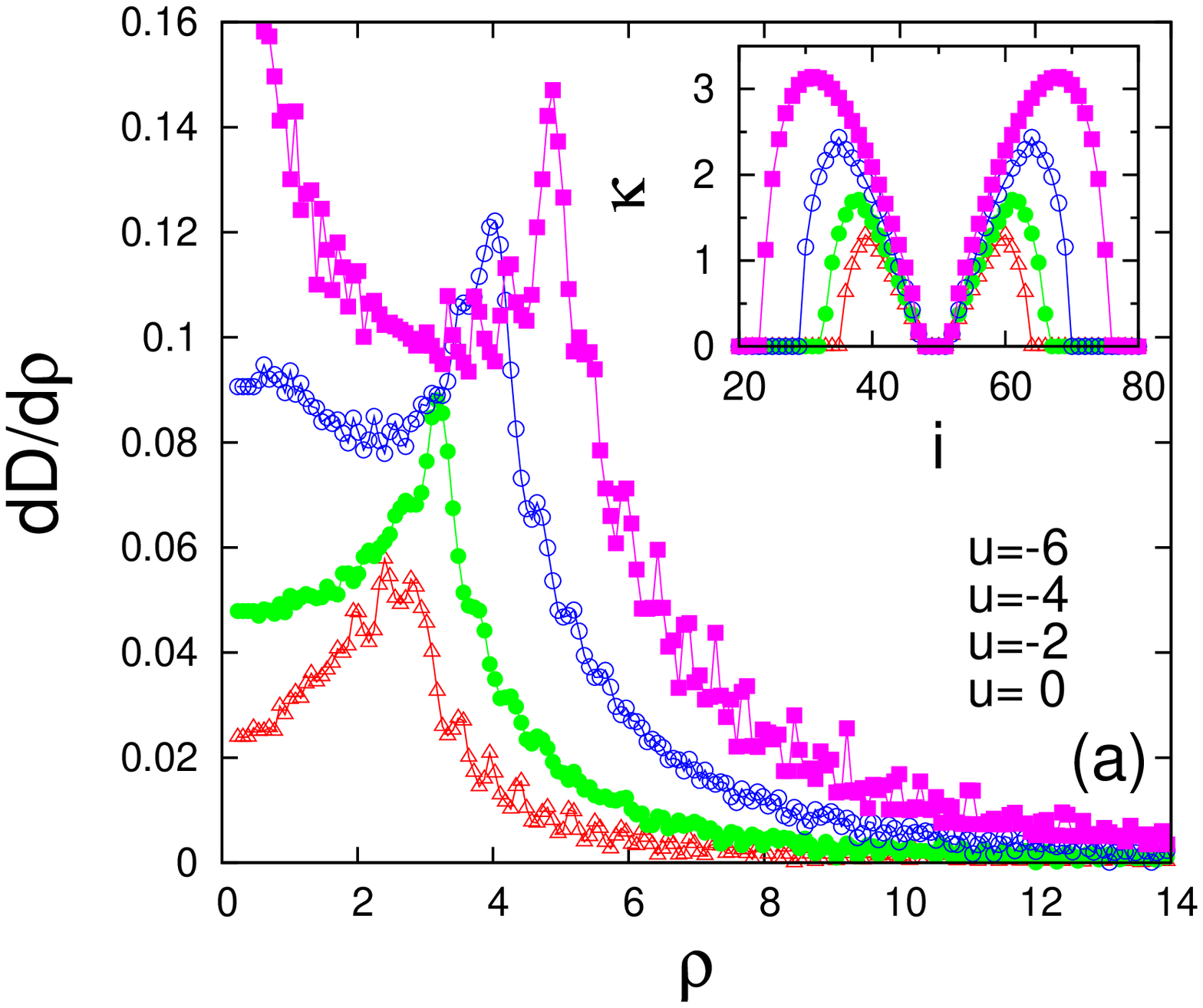}\\
\includegraphics*[width=0.9\linewidth]{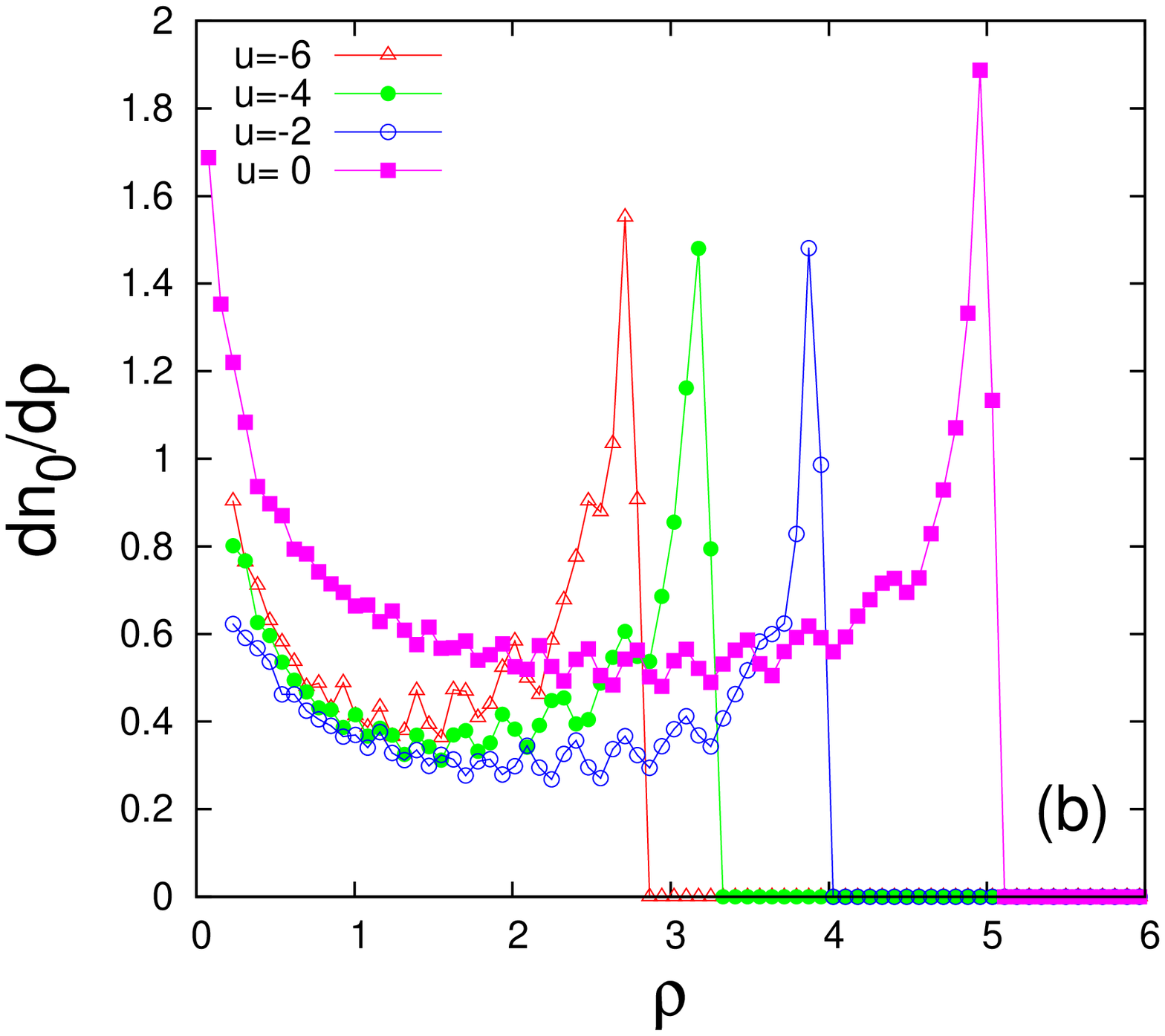}
\caption{(Color online) (a) The derivative of the ground-state concentration of doubly occupied sites $D$ as a function of characteristic density $\rho$ in the system under the presence of trapping potential. In the inset, the local inverse compressibility $\kappa^{-1}_i$ (in units of $t$) is shown as a function of $i$ with the system parameters determined from the critical characteristic density on the peaks. (b) The derivative of the local central density $n_0$ as a function of characteristic density $\rho$ for different attractive interactions.
For comparison, the derivatives of the double occupancy and the local central density for the noninteracting case ($u=0$) are also shown in (a) and (b), respectively.
\label{fig:four}}
\end{center}
\end{figure}

The harmonic trapping potential in Eq. (\ref{eq:hubbard}) induces a position dependence of the chemical potentials which can be evaluated within the local density approximation based on the Bethe-ansatz equations. This amounts to determining the global chemical potential $\mu$ of the system from the local equilibrium condition,
\begin{equation}\label{eq:gmu}
\mu=\left.\mu_{\rm hom}(n,u)\right|_{n\rightarrow n_i}+V_2 (i-N_{s}/2)^2,
\end{equation}
while imposing the normalization condition $\sum_i n_i=N_f$.
Here $\mu_{\rm hom}(n,u)$ is the chemical potential of the homogeneous system calculated from the Bethe-ansatz coupled equations~\cite{gao_long_2005,gaoprb78}.

In Fig. 1, we show the above-mentioned composite phases through the $\rho-u$ phase diagram, which is insensitive to the individual values of $N_f$ and $V_2$, that is, for fixed $u$, the shape of the density profile in this system is completely determined by $\rho$~\cite{Rigol}. Below the BI phase, it is the LE liquid phase. We would like to mention that the similar phase diagram has been obtained by Rigol et al. for the repulsive interaction~\cite{Rigol} and Heiselberg but with a fitting function for $\mu_{\rm hom}(n,u)$~\cite{Heiselberg}. Here we solve the local equilibrium condition with the exact numerical data for $\mu_{\rm hom}(n,u)$ calculated from the coupled Bethe-ansatz equations. We conclude that the phase diagram based on the Thomas-Fermi approximation gives qualitatively correct results when comparing to QMC.

To detect the composite phases induced by the trapping potential, we investigate firstly the concentration of the doubly occupied sites, which is defined as the derivative of the ground state energy with respect to $U$~\cite{Leo},
\[
D=\frac{2}{N_f}\sum^{N_s}_{i=1}\langle \hat{D}_i\rangle
=\frac{2}{N_f}\frac{\partial E_0}{\partial U}\,.
\]
Here $E_0$ is the ground state energy per site of the system.
The DO in the homogeneous system, $D^{\rm hom}$, can be calculated exactly from the Bethe-ansatz coupled equations~\cite{KocharianPRB}, which is shown in Fig. 2. Notice, at $U=0$, we have $D^{\rm hom}=n^2/4$ and at strongly attractive coupling limit $D^{\rm hom}= n/2$. In the presence of trapping potential, we calculate the DO in the spirit of local density approximation, $D=\frac{2}{N_f}\sum_i D^{\rm hom}(n)|_{n\rightarrow n_i}$. From Fig. 2, it becomes obvious that the attractive interaction favors the formation of pairs between the two spin species while the repulsive interaction suppresses them.

Now we show in Fig. 3 (a) the DO and in Fig. 3 (b) the local central density $n_0$ for the inhomogeneous system described in Eq. (1) as a function of the characteristic density $\rho$. For large $\rho$, the DO and the central density saturate at $D\approx 1$ and $n_0\approx 2$, respectively, corresponding to the BI regime, where an incompressible phase of band-insulating type is realized in the core of the trap. The more attractive of the interaction is, the sooner for $D$ to reach the value of $1$ and for $n_0$ to reach $2$. These manifest that the attractive interaction favors pairing, and thus increases DO, different from the case of repulsive interaction where it suppresses DO and supports Mott-insulating phase.

In the following, we show the derivative of the DO in Fig. 4 (a) and the derivative of the local central density in Fig. 4 (b), respectively, as probes for the composite quantum phase transitions and especially as signs for band-insulating phases. The derivatives of the DO and the local central density are performed numerically using the central finite difference method.
We find that the singularities in these two quantities provide a further support and clear signatures of the composite quantum phase transitions accompanying Fig. 3. In Fig. 4 (a), the phase transition from LE to BI is manifested by an obvious peak. The onset of the peak gives a critical value of the characteristic density, which determines the critical particle number for a given $V_2/t$. Based on these parameters, we plot in the inset of Fig. 4 (a) the local inverse compressibility defined by $\kappa^{-1}_i=\partial^2 E_{0}/\partial n^2|_{n\rightarrow n_i}$ as a function of site position. The incompressible nature of the BI phase is manifested by the disappearance of $\kappa^{-1}_i$, which is compatible with the onset of the peak in $dD/d\rho$. In Fig. 4 (b), the vanishing of the derivative of the local central density gives very direct signatures of the incompressible nature of BI in the center of the trap. The phase boundaries of the onset of the BI phase are plotted in Fig. \ref{fig:one}, which are determined by the appearance of the peak in the derivative of the DO and the vanishing of the derivative of the local central density. We find that the phase boundary determined from the DO slightly overestimates the onset of the BI phase, while the central density slightly underestimates it.

In conclusion, motivated by earlier theoretical work and by ongoing experimental efforts, we present a fully numerical study for the ground state properties of a one-dimensional attractive Fermion-Hubbard model in harmonic traps within the Thomas-Fermi approximation based on the exact Bethe-ansatz solution. We investigate the double occupancy, the local central density and their derivatives. We found they are good indicators of the formation of an incompressible band-insulating phase in the center of the trap, which can be detected in the near-future in the experiments in the 1D cold fermionic atomic gases.

This work was supported by NSF of China under Grant No. 10704066, 10974181, Qianjiang River Fellow Fund 2008R10029, and Program for Innovative Research Team in Zhejiang Normal University.

\end{document}